\documentclass[runningheads]{llncs}
\usepackage{graphicx}
\usepackage{textcomp}
\usepackage{multirow}
\usepackage{booktabs}

\begin{document}
\title{Photoacoustic Microscopy with Sparse Data Enabled by Convolutional Neural Networks for Fast Imaging}
%
%
\author{Jiasheng Zhou\inst{1}\inst{*} \and Da He\inst{1}\inst{*} \and Xiaoyu Shang\inst{1} \and Zhendong Guo\inst{1} \and Sung-liang Chen\inst{1}\inst{\dag} \and Jiajia Luo\inst{2}\inst{\dag}}
%
\authorrunning{J. Zhou et al.}
%
\institute{University of Michigan-Shanghai Jiao Tong University Joint Institute, Shanghai Jiao Tong University, Shanghai 200240, China\\
\email{sungliang.chen@sjtu.edu.cn} \and
Biomedical Engineering Department, Peking University, Beijing 100191, China\\
\email{jiajia.luo@pku.edu.cn}
}
%
\maketitle              
\footnote{$^*$ Contributed equally.\\ $^{\dag}$ Corresponding authors.}
\begin{abstract}
	Photoacoustic microscopy (PAM) has been a promising biomedical imaging technology in recent years. However, the point-by-point scanning mechanism results in low-speed imaging, which limits the application of PAM. Reducing sampling density can naturally shorten image acquisition time, which is at the cost of image quality. In this work, we propose a method using convolutional neural networks (CNNs) to improve the quality of sparse PAM images, thereby speeding up image acquisition while keeping good image quality. The CNN model utilizes both squeeze-and-excitation blocks and residual blocks to achieve the enhancement, which is a mapping from a 1/4 or 1/16 low-sampling sparse PAM image to a latent fully-sampled image. The perceptual loss function is applied to keep the fidelity of images. The model is mainly trained and validated on PAM images of leaf veins. The experiments show the effectiveness of our proposed method, which significantly outperforms existing methods quantitatively and qualitatively. Our model is also tested using {\it in vivo} PAM images of blood vessels of mouse ears and eyes. The results show that the model can enhance the image quality of the sparse PAM image of blood vessels from several aspects, which may help fast PAM and facilitate its clinical applications.

\keywords{Photoacoustic microscopy  \and Convolutional neural network \and  Sparse image.}
\end{abstract}
\section{Introduction}
%
Photoacoustic microscopy (PAM), as a hybrid imaging technique based on the photoacoustic (PA) effect~\cite{b1,b2}, has been widely used for biomedical imaging~\cite{b3,b4,b5,b6}. Optical-resolution PAM (OR-PAM), as one implementation of PAM, offers high spatial resolution at the expense of penetration depth and has demonstrated many potential applications~\cite{b6,b7,b8}. For image acquisition in OR-PAM, since a sample typically has spatially different optical absorption, the absorption map of the sample is obtained by point-by-point scanning over the sample. As a result, the imaging speed of OR-PAM is highly restricted by the point-by-point scanning mechanism, especially for high-resolution OR-PAM that performs scanning with small step size and thus more scanning points (i.e., more pixels) within a specific region of interest (ROI). A low imaging frame rate may hamper selected applications such as monitoring dynamic biological systems.

In recent years, efforts to increase the scanning speed of OR-PAM have mainly focused on the fast scanning mechanism. Components for fast-scanning PAM include high-speed voice-coil stages~\cite{b9,b10}, galvanometer scanners~\cite{b11}, microelectromechanical system scanning mirrors~\cite{b12}, and micro lens arrays and array ultrasonic transducers~\cite{b13,b14}. A random-access scanning method is also applied in OR-PAM to improve imaging speed by scanning a selected region using a digital micromirror device~\cite{b15}. In these works, sophisticated and expensive hardware is used. Instead, sparse-scanning OR-PAM offers an alternative solution that saves image acquisition time by reducing scanning points (compared to full scan) to increase imaging speed. Methods such as compressive sensing are also applied to recover a PAM image with sparse data~\cite{b16}. However, due to the relatively high sampling density and the complexity of compression sampling experiments, their contribution to imaging speed is limited. Therefore, there is still a need for more efficient and practical algorithms to accelerate the imaging speed of OR-PAM.

Convolutional neural networks (CNNs), as a promising method, have been used to achieve super resolution (SR)~\cite{b17,b18,b19} and enhance the PA computed tomography~\cite{b20,b21,b22,b23,b24}. However, to our knowledge, there are no studies to improve the PAM imaging speed using CNN-based methods. Enhancing the quality of a low-sampling sparse PAM image ( i.e., restoring it to a latent fully-sampled image), can be categorized as an SR problem (i.e., low-resolution image to high-resolution image). Therefore, a CNN-based method can be utilized to improve the quality of the sparse PAM image. In this work, we propose to use CNNs to process sparse PAM images, so sparse scanning can be used to increase the imaging speed. High-quality images are restored from 1/4 or 1/16 low-sampling images using the proposed CNN model. The model is trained and validated on a dataset consisting of 268 PAM images of leaf veins, which can be accessed online for further studies by other researchers. We also extend our method to {\it in vivo} applications and achieve high performance in restoring sparse PAM images of blood vessels of mouse ears and eyes, demonstrating the feasibility in biomedical research and clinical application.

\section{Methods}
\subsection{Dataset Preparation}
In this work, a dataset of PAM images of oak and magnolia leaf veins was used to train and validate our CNN model. Leaves were immersed in a container with black ink for more than 7 hours. Then, the leaves were placed on a glass slide and sealed with silicone glue (GE Sealants). For each PAM image, the leaf samples were scanned by an OR-PAM probe (resolution: 2 \textmu m, measured by a beam profiler and a $10\times$ beam expander) over $256\times256$ scanning points with scanning step size of 8 \textmu m. Finally, we acquired a dataset of 268 original full-sampling PAM images in total using our PAM system (see Fig. S1 in Supplementary Material).

\begin{figure}
	\includegraphics[width=\textwidth]{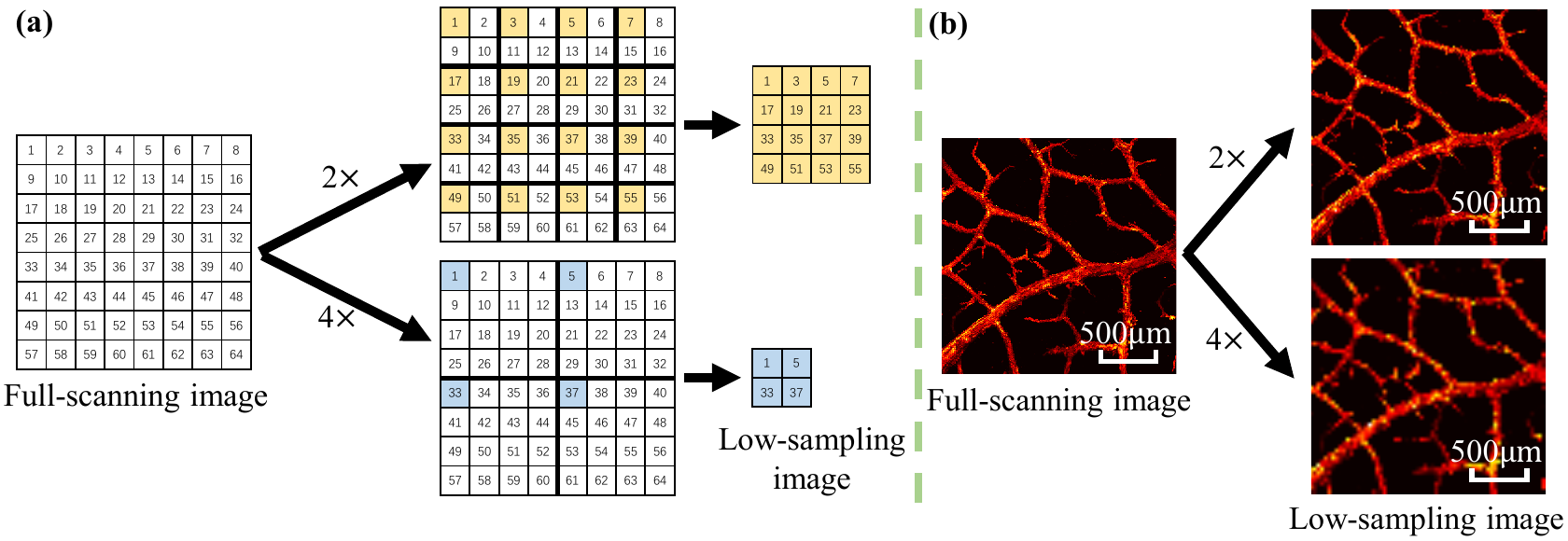}
	\caption{The process of generating low-sampling images from the full-sampling (i.e., full-scanning) ones. (a) Illustration of the down-sampling method. (b) An example by applying the down-sampling method.} \label{fig1}
\end{figure}

When generating low-sampling sparse data from full-sampling data, as shown in Fig.~\ref{fig1}, with $2\times$ scaling in step size, only 1/2 pixels in one lateral dimension are selected and used (as indicated by the yellow-colored pixels in Fig.~\ref{fig1}(a)) in the low-sampling image. That is, the low-sampling image ($128\times128$ pixels) has only 1/4 pixels of the full-sampling image. Similarly, with a $4\times$ step size scaling, the low-sampling image ($64\times64$ pixels) has only 1/16 pixels of the full-sampling case. It is also expected to require only 1/16 image acquisition time of the full-sampling image. Finally, 268 pieces of raw data are collected for our CNN model, where the 2D low-sampling PAM image is used as input and 2D full-sampling PAM image ($256\times256$ pixels) is used as output (i.e., ground truth). Note that the 2D PAM image here means 2D maximum amplitude projection (MAP) along the depth direction, which is commonly used for the OR-PAM image display. As can be seen, the image quality is degraded in the low-sampling images of Fig.~\ref{fig1}(b) (e.g., blurs and discontinuities). For each scaling rate ($2\times$ and $4\times$), we split the dataset into training, validation, and test sets with a ratio of 0.8:0.1:0.1. Regular data augmentation operations, including flipping and rotation, are applied for training.

\subsection{Network Architecture and Settings}
\begin{figure}[t]
	\includegraphics[width=\textwidth]{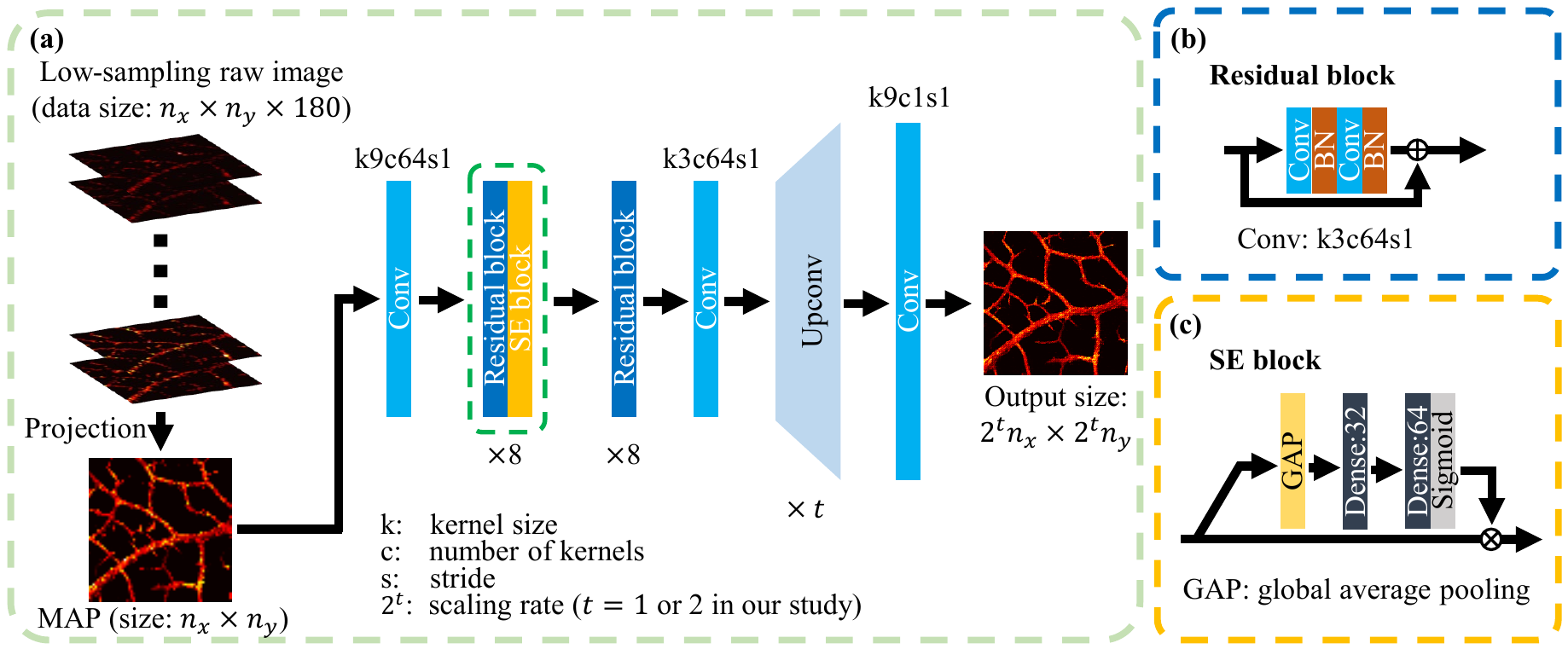}
	\caption{ The architecture of the proposed CNN model. (a) The overview of the proposed CNN model. (b) The details of each residual block. (c) The details of each SE block.} \label{fig2}
\end{figure}

The architecture of the proposed CNN is shown in Fig.~\ref{fig2}. We utilize 16 residual blocks and 8 Squeeze-and-Excitation (SE) blocks~\cite{b25} as the key parts of feature extraction. Inspired by SRGAN~\cite{b19}, the residual blocks elaborated in Fig.~\ref{fig2}(b) can well extract features in SR tasks. Moreover, we find that the SE block~\cite{b25} (shown in Fig.~\ref{fig2}(c)) with the channel-wise attention mechanism contributes to network convergence and performance. The ``Upconv'' block consists of a $2\times$ upsampling layer and a standard convolutional layer (with a kernel size of 3, number of filters of 256, and stride of 1). The final output layer is followed by a Tanh activation function.

The perceptual loss is applied to train the CNN model. As indicated in~\cite{b19,b26,b27,b28,b29}, although pixel-wise mean squared error (MSE) loss function significantly improves the pixel-wise metrics such as peak signal-to-noise ratio (PSNR) and structural similarity index (SSIM)~\cite{b30}, the generated image is easily too smooth and the quality is poor from a subjective point of view. This phenomenon is quite severe in our PAM images (demonstrated later). Instead, for the perceptual loss, we calculate the MSE based on the output feature map of the 7th convolutional layer of VGG19~\cite{b19,b26,b29,b31,b32}, which can give the high-level feature description of the image. The VGG19 model is pretrained on the ImageNet dataset~\cite{b33}. With the prediction $I^{pred}$ and the ground truth projection $I^{gt}$ to calculate the perceptual loss, the two feature maps from the 7th convolutional layer of VGG19 can be expressed as $f(I^{pred})$ and $f(I^{gt})$, respectively. Thus, the perceptual MSE loss should be:

\begin{equation}Loss\!=\!\frac{1}{n_x\!\times\! n_y\!\times\! n_z}\sum_{x=1}^{n_x}\sum_{y=1}^{n_y}\sum_{z=1}^{n_z}(f(I^{gt})_{x,y,z} \!-\! f(I^{pred})_{x,y,z})^2 ,\label{eq}\end{equation}
where $n_x$, $n_y$, and $n_z$ denote the dimensions of the feature map.

In our experiments, the proposed CNN model is implemented using Keras framework with Tensorflow backend. Adam optimizer is applied with $\beta_1=0.5$. The learning rate is 2e-4. A single Nvidia 2080Ti GPU is used for training.

\section{Results and Analysis}
\subsection{Leaf Vein Experiment by the Down-sampling Method for Testing}
\begin{figure}[t]
	\includegraphics[width=\textwidth]{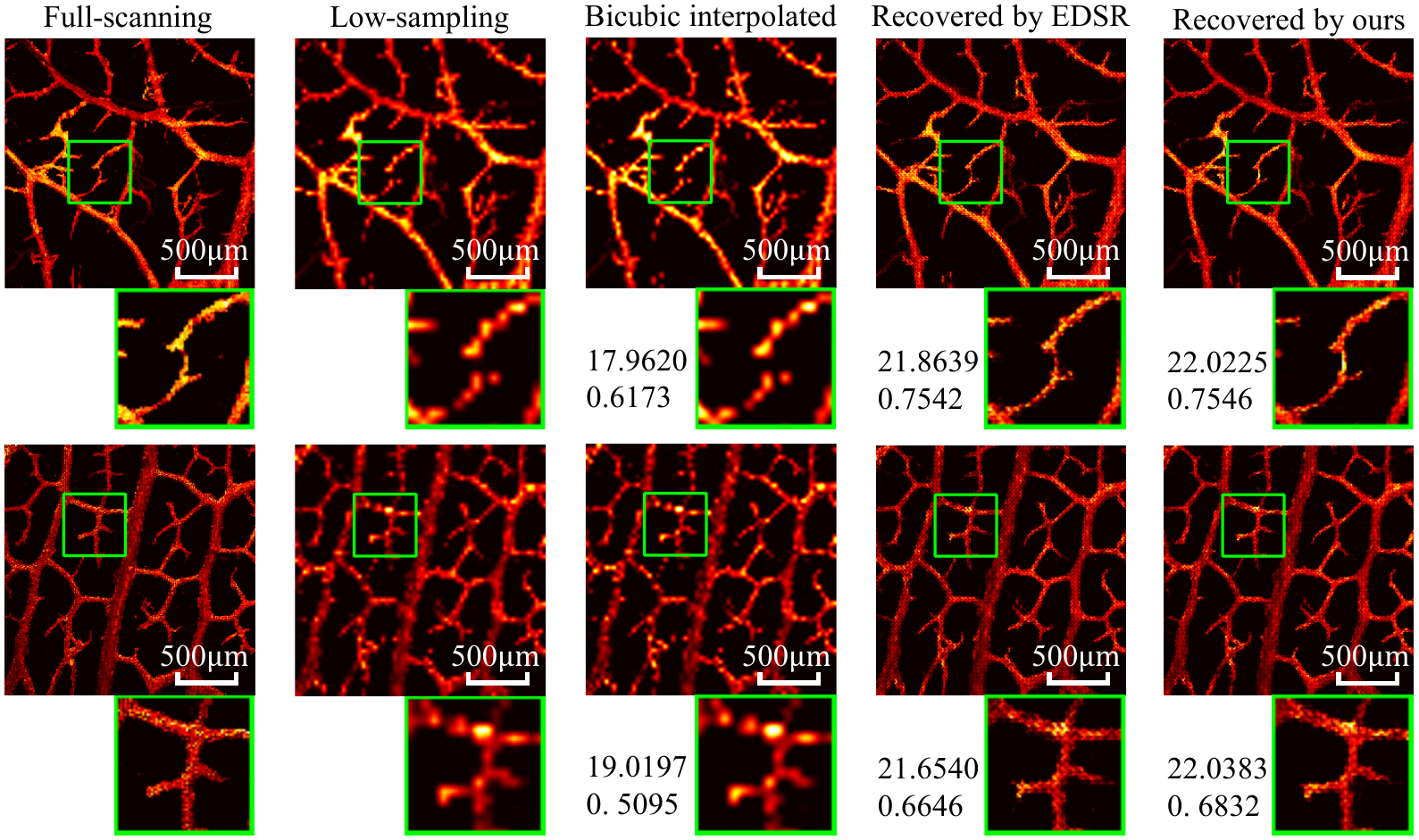}
	\caption{Example results of the leaf vein experiment. The numbers below images indicate the PSNR (dB) and SSIM values compared with the corresponding ground truth. The first sample is from a magnolia leaf and the second is from an oak leaf.} \label{fig3}
\end{figure}

Fig.~\ref{fig3} shows two representative results of image restoration for $4\times$ scaling rate (results for $2\times$ scaling rate can be found in Supplementary Material). Besides our restoration method, two other representative methods, the bicubic interpolation and a re-trained EDSR model~\cite{b34} are applied for comparison. EDSR is a typical and effective CNN-based method originally designed for natural images.

By checking zoomed images (denoted by the green boxes), the two CNN-based methods are superior to bicubic interpolation. Specifically, first, the results by bicubic interpolation were blurred and overly smoothed. Secondly, the low-sampling image suffered from discontinuities, which were not recovered by bicubic. By contrast, no such issues are observed by CNN methods, and the recovered images look more natural and closer to the full-scanning ones. Bicubic interpolation (and other conventional methods) uses the weighted average values of a local area, while the CNN model can learn more high-level (or more global) information to predict pixel values better. It is worth noting that the above issues (over smoothing, blurring, and discontinuity) become more severe in the recovered images using bicubic interpolation from 1/4 to 1/16 low-sampling cases, while the quality of the recovered images is surprisingly maintained almost the same from 1/4 to 1/16 low-sampling cases using our CNN method (see Supplementary Material for 1/4 low-sampling cases). That is, as the scaling increases, the advantages of our method becomes more apparent than the bicubic.

A similar trend can be found in the statistical results on the test set, as shown in Table~\ref{table:1}. For the $4\times$ scaling case, compared to bicubic interpolation, our model's PSNR and SSIM values (average of the test set) are greatly improved by 3.1819 dB and 0.1386. Besides, according to the two metrics, our model outperforms the re-trained EDSR model at both $2\times$ and $4\times$ scaling.
\begin{table}
	\centering
	\caption{Leaf vein experiment: Comparison of PSNR and SSIM values.}
	\label{table:1}
	\begin{tabular}[c]{lcccc}
		\toprule
		\multirow{2}{*}{  }&
		\multicolumn{2}{c}{$2\times$} & \multicolumn{2}{c}{$4\times$} \\
		\cmidrule(lr){2-5}
		& PSNR (dB) & SSIM & PSNR (dB) & SSIM \\
		\midrule
		Bicubic & 23.4936 & 0.7721 & 19.9941 & 0.5773\\
		EDSR & 24.2356 & 0.5955 & 21.5557 & 0.6264\\
		Ours & \textbf{26.1431} & \textbf{0.8183} & \textbf{23.1760} & \textbf{0.7159}\\
		\bottomrule
	\end{tabular}
\end{table}

\subsection{Leaf Vein Experiment Using Experimentally-acquired Sparse Data for Verification}

To further verify the feasibility of our method, besides the low-sampling images obtained from the operation in Fig.~\ref{fig1}, experimentally-acquired sparse PAM images are also fed to our trained CNN, which is closer to practical applications. We scan the same ROI with a scanning step size of 8 \textmu m and 16 \textmu m (or 32 \textmu m). That is, the PAM image pair with lateral sizes of $256\times256$ and $128\times128$ (or $64\times64$)pixels is experimentally scanned for the same ROI. The low-sampling PAM images are used as the input and the corresponding image with $256\times256$ pixels is used as the reference. One representative result is shown in Supplementary Material, where the advantages by our CNN model (no issues of over smoothing, blurring, and discontinuity) are also observed. The results verify that by using our CNN model for sparse-scanning and post processing, fast PAM imaging can be realized to achieve images with similar quality to the very time-consuming full-sampling corresponding image.

\subsection{Ablation Investigation}
As shown in Fig.~\ref{fig2}, we applied the SE block~\cite{b25} after some residual blocks. With the channel-wise attention design, SE block is thought to be useful for channel information selection. In our experiments, the CNN without SE blocks shows relatively poor results. For example, for $4\times$ scaling test set, SE blocks can improve the PSNR and SSIM values by 1.3876 dB and 0.0936, respectively. The detailed comparison results can be found in Supplementary Material. 

\begin{figure}[tb]
	\centering
	\includegraphics[width=0.9\textwidth]{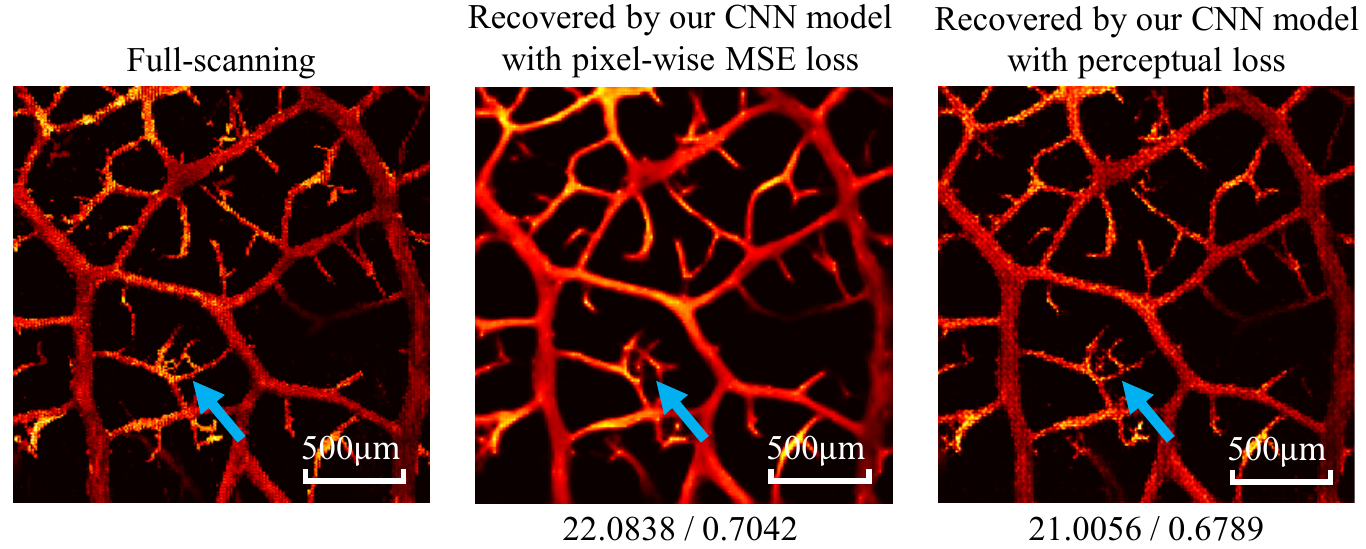}
	\caption{Example results of CNN models with pixel-wise MSE loss and perceptual loss. The numbers below restored images indicate the corresponding PSNR (dB) / SSIM values. The sample comes from an oak leaf.} \label{fig4}
\end{figure}

The perceptual loss is one of the most critical settings of our method. As explained before, training with pixel-wise MSE probably results in finding a pixel-wise average solution, which loses fine texture~\cite{b19,b26,b27,b28,b29}. To illustrate the problem, we have trained a pair of models: one used the perceptual loss function while the other utilized the standard pixel-wise MSE loss function. Example results for the $4\times$ scaling case are shown in Fig.~\ref{fig4}. According to Fig.~\ref{fig4}, PSNR and SSIM values of the middle image are higher than those of the right one. However, the middle image is so smooth that it differs from the ground truth (i.e., full-sampling) a lot from the perceptive point of view. Some small branches in the middle PAM image even disappear (e.g., the parts indicated by the blue arrows in Fig.~\ref{fig4}). By contrast, the right PAM image looks very much like the corresponding ground truth (e.g., restored more textures in the ground truth), which may be more critical to biomedical applications. In this regard, it is essential to apply such a perceptual loss function.

\subsection{{\it In vivo} Experiment}

\begin{figure}[t]
	\centering
	\includegraphics[width=\textwidth]{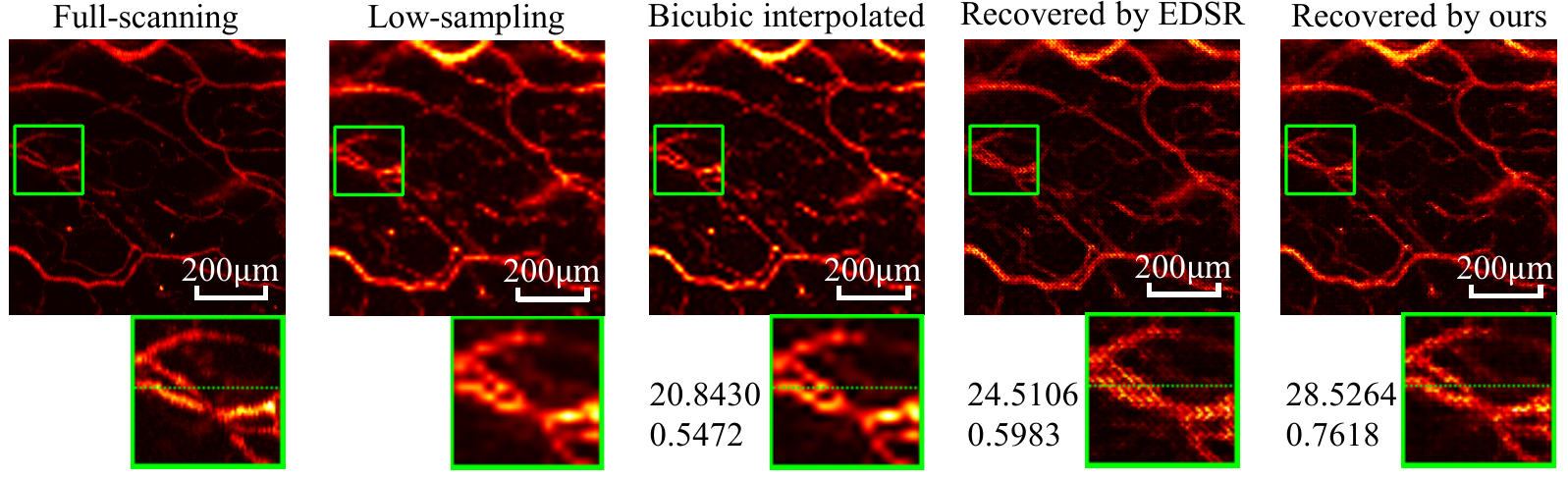}
	\caption{Demonstration of {\it in vivo} PAM images of blood vessels of the mouse ear. The numbers below images show the PSNR (dB) and SSIM values compared with the corresponding ground truth. 1D profiles along the dashed green lines in the zoom images are attached in Supplementary Material.} \label{fig5}
\end{figure}

Without transfer learning, our model is only trained with the leaf veins dataset and then directly used to test the PAM images of mouse ear and eye, showing improvements quantitatively and qualitatively. A representative sample for $4\times$ scaling case is shown in Fig.~\ref{fig5}. The full-sampling PAM image is acquired using the probe with a resolution of 4 \textmu m and a scanning step size of 3 \textmu m. As can be seen, the recovered PAM images using our method show sharper edges and more distinguishable patterns compared with those by bicubic interpolation and EDSR. For better comparison, one-dimensional (1D) profiles in Fig.~\ref{fig5} are plotted in Supplementary Material.

\begin{figure}[h]
	\includegraphics[width=\textwidth]{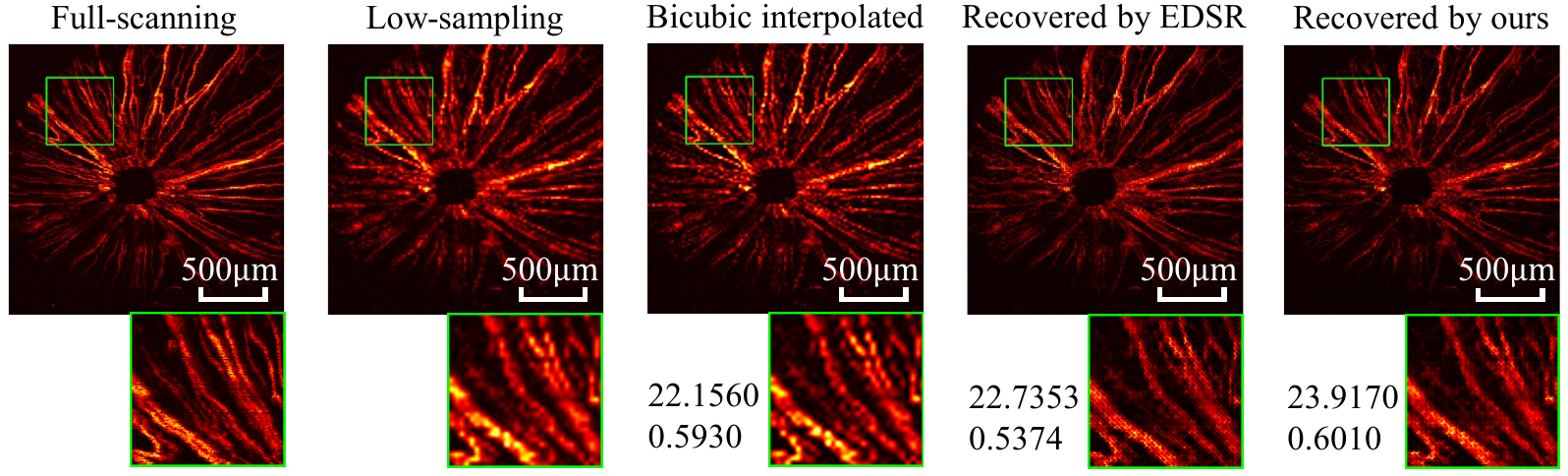}
	\caption{Demonstration of {\it in vivo} PAM images of blood vessels of the mouse eye. The numbers below images show the PSNR (dB) and SSIM values compared with the corresponding ground truth.} \label{fig6}
\end{figure}

Further, we attempt to test our model for different patterns other than tree-like patterns (i.e., with branches, subbranches, etc.) demonstrated previously. Therefore, an {\it in vivo} PAM image of blood vessels of the mouse eyes with radial patterns was tested. The raw data were acquired by the probe with a resolution of $\sim$3 \textmu m and a scanning step size of 4 \textmu m~\cite{b35}. Fig.~\ref{fig6} shows the results for the $4\times$ scaling case. Our method renders a better-recovered PAM image than bicubic interpolation and EDSR. This can be appreciated more clearly by comparing the zoom images. Therefore, even if the CNN model trained from tree-like pattern images is applied to an image with radial patterns, we still achieve good performance, thereby showing the robustness of the CNN method to some degree.

\section{Conclusion}
We propose a novel CNN-based method to improve the quality of sparse PAM images, which can equivalently improve PAM imaging speed. The model is trained on the dataset of PAM images of leaf samples. Residual blocks, SE bocks, and perceptual loss function are essential in our CNN model. Both 1/4 and 1/16 low-sampling sparse PAM images (i.e., $2\times$ and $4\times$ scaling cases, respectively) were tested, and the proposed CNN method showed remarkable performance both quantitatively and intuitively. We have also tested our method using {\it in vivo} PAM images of blood vessels of mouse ears and eyes, and the recovered PAM images had a high resemblance to the full-sampling ones. The CNN method to deal with sparse data demonstrated in OR-PAM may also be applied to AR-PAM and other point-by-point scanning imaging modalities such as optical coherence tomography and confocal fluorescence microscopy. Our work opens up new opportunities for fast PAM imaging.

\bibliographystyle{splncs04}
\bibliography{refs}

\begin{thebibliography}{10}
\providecommand{\url}[1]{\texttt{#1}}
\providecommand{\urlprefix}{URL }
\providecommand{\doi}[1]{https://doi.org/#1}

\bibitem{b22}
Allman, D., Reiter, A., Bell, M.A.L.: Photoacoustic source detection and
  reflection artifact removal enabled by deep learning. IEEE Trans. Med.
  Imaging  \textbf{37}(6),  1464--1477 (2018)

\bibitem{b21}
Anas, E.M.A., Zhang, H.K., Kang, J., Boctor, E.: Enabling fast and high quality
  led photoacoustic imaging: a recurrent neural networks based approach.
  Biomed. Opt. Express  \textbf{9}(8),  3852--3866 (2018)

\bibitem{b23}
Antholzer, S., Haltmeier, M., Nuster, R., Schwab, J.: Photoacoustic image
  reconstruction via deep learning. In: Photons Plus Ultrasound: Imaging and
  Sensing 2018. vol. 10494, p. 104944U. International Society for Optics and
  Photonics (2018)

\bibitem{b24}
Antholzer, S., Haltmeier, M., Schwab, J.: Deep learning for photoacoustic
  tomography from sparse data. Inverse. Probl. Sci. En.  \textbf{27}(7),
  987--1005 (2019)

\bibitem{b3}
Beard, P.: Biomedical photoacoustic imaging. Interface Focus  \textbf{1}(4),
  602--631 (2011)

\bibitem{b1}
Bell, A.G.: Art. xxxiv.--on the production and reproduction of sound by light.
  Am. J. Sci. (1880-1910)  \textbf{20}(118), ~305 (1880)

\bibitem{b26}
Bruna, J., Sprechmann, P., LeCun, Y.: Super-resolution with deep convolutional
  sufficient statistics. arXiv preprint arXiv:1511.05666  (2015)

\bibitem{b33}
Deng, J., Dong, W., Socher, R., Li, L.J., Li, K., Fei-Fei, L.: Imagenet: A
  large-scale hierarchical image database. In: Proceedings of the IEEE
  Conference on Computer Vision and Pattern Recognition. pp. 248--255 (2009)

\bibitem{b18}
Dong, C., Loy, C.C., He, K., Tang, X.: Image super-resolution using deep
  convolutional networks. IEEE Trans. Pattern Anal. Mach. Intell.
  \textbf{38}(2),  295--307 (2015)

\bibitem{b28}
Dosovitskiy, A., Brox, T.: Generating images with perceptual similarity metrics
  based on deep networks. In: Advances in Neural Information Processing
  Systems. pp. 658--666 (2016)

\bibitem{b32}
Gatys, L., Ecker, A.S., Bethge, M.: Texture synthesis using convolutional
  neural networks. In: Advances in Neural Information Processing Systems. pp.
  262--270 (2015)

\bibitem{b17}
Goodfellow, I., Pouget-Abadie, J., Mirza, M., Xu, B., Warde-Farley, D., Ozair,
  S., Courville, A., Bengio, Y.: Generative adversarial nets. In: Advances in
  Neural Information Processing Systems. pp. 2672--2680 (2014)

\bibitem{b35}
Guo, Z., Li, Y., Chen, S.L.: Miniature probe for in vivo optical-and
  acoustic-resolution photoacoustic microscopy. Opt. Lett.  \textbf{43}(5),
  1119--1122 (2018)

\bibitem{b11}
Hajireza, P., Shi, W., Bell, K., Paproski, R.J., Zemp, R.J.:
  Non-interferometric photoacoustic remote sensing microscopy. Light: Sci.
  Appl.  \textbf{6}(6),  e16278--e16278 (2017)

\bibitem{b9}
Harrison, T., Ranasinghesagara, J.C., Lu, H., Mathewson, K., Walsh, A., Zemp,
  R.J.: Combined photoacoustic and ultrasound biomicroscopy. Opt. Express
  \textbf{17}(24),  22041--22046 (2009)

\bibitem{b25}
Hu, J., Shen, L., Sun, G.: Squeeze-and-excitation networks. In: Proceedings of
  the IEEE Conference on Computer Vision and Pattern Recognition. pp.
  7132--7141 (2018)

\bibitem{b14}
Imai, T., Shi, J., Wong, T.T., Li, L., Zhu, L., Wang, L.V.: High-throughput
  ultraviolet photoacoustic microscopy with multifocal excitation. J. Biomed.
  Opt.  \textbf{23}(3),  036007 (2018)

\bibitem{b6}
Jeon, S., Kim, J., Lee, D., Woo, B.J., Kim, C.: Review on practical
  photoacoustic microscopy. Photoacoustics p. 100141 (2019)

\bibitem{b29}
Johnson, J., Alahi, A., Fei-Fei, L.: Perceptual losses for real-time style
  transfer and super-resolution. In: Proceedings of the European Conference on
  Computer Vision. pp. 694--711 (2016)

\bibitem{b19}
Ledig, C., Theis, L., Husz{\'a}r, F., Caballero, J., Cunningham, A., Acosta,
  A., Aitken, A., Tejani, A., Totz, J., Wang, Z., et~al.: Photo-realistic
  single image super-resolution using a generative adversarial network. In:
  Proceedings of the IEEE Conference on Computer Vision and Pattern
  Recognition. pp. 4681--4690 (2017)

\bibitem{b13}
Li, G., Maslov, K.I., Wang, L.V.: Reflection-mode multifocal optical-resolution
  photoacoustic microscopy. J. Biomed. Opt.  \textbf{18}(3),  030501 (2013)

\bibitem{b15}
Liang, J., Zhou, Y., Winkler, A.W., Wang, L., Maslov, K.I., Li, C., Wang, L.V.:
  Random-access optical-resolution photoacoustic microscopy using a digital
  micromirror device. Opt. Lett.  \textbf{38}(15),  2683--2686 (2013)

\bibitem{b34}
Lim, B., Son, S., Kim, H., Nah, S., Mu~Lee, K.: Enhanced deep residual networks
  for single image super-resolution. In: Proceedings of the IEEE conference on
  Computer Vision and pattern recognition workshops. pp. 136--144 (2017)

\bibitem{b16}
Liu, T., Sun, M., Liu, Y., Hu, D., Ma, Y., Ma, L., Feng, N.: Admm based
  low-rank and sparse matrix recovery method for sparse photoacoustic
  microscopy. Biomed. Signal Process.  \textbf{52},  14--22 (2019)

\bibitem{b8}
Liu, W., Yao, J.: Photoacoustic microscopy: principles and biomedical
  applications. Biomed. Eng. Lett.  \textbf{8}(2),  203--213 (2018)

\bibitem{b27}
Mathieu, M., Couprie, C., LeCun, Y.: Deep multi-scale video prediction beyond
  mean square error. arXiv preprint arXiv:1511.05440  (2015)

\bibitem{b20}
Ronneberger, O., Fischer, P., Brox, T.: U-net: Convolutional networks for
  biomedical image segmentation. In: Navab, N., Hornegger, J., Wells, W.,
  Frangi, A. (eds.) MICCAI 2015. vol.~9351, pp. 234--241. Springer, Cham (2015)

\bibitem{b31}
Simonyan, K., Zisserman, A.: Very deep convolutional networks for large-scale
  image recognition. arXiv preprint arXiv:1409.1556  (2014)

\bibitem{b10}
Wang, L., Maslov, K.I., Xing, W., Garcia-Uribe, A., Wang, L.V.: Video-rate
  functional photoacoustic microscopy at depths. J. Biomed. Opt.
  \textbf{17}(10),  106007 (2012)

\bibitem{b4}
Wang, L.V., Hu, S.: Photoacoustic tomography: in vivo imaging from organelles
  to organs. Science  \textbf{335}(6075),  1458--1462 (2012)

\bibitem{b7}
Wang, L.V., Yao, J.: A practical guide to photoacoustic tomography in the life
  sciences. Nat. Methods  \textbf{13}(8), ~627 (2016)

\bibitem{b30}
Wang, Z., Bovik, A.C., Sheikh, H.R., Simoncelli, E.P.: Image quality
  assessment: from error visibility to structural similarity. IEEE Trans. Image
  Process.  \textbf{13}(4),  600--612 (2004)

\bibitem{b12}
Yao, J., Wang, L., Yang, J.M., Maslov, K.I., Wong, T.T., Li, L., Huang, C.H.,
  Zou, J., Wang, L.V.: High-speed label-free functional photoacoustic
  microscopy of mouse brain in action. Nat. Methods  \textbf{12}(5),  407--410
  (2015)

\bibitem{b2}
Yao, J., Wang, L.V.: Photoacoustic microscopy. Laser Photonics Rev.
  \textbf{7}(5),  758--778 (2013)

\bibitem{b5}
Yao, J., Wang, L.V.: Recent progress in photoacoustic molecular imaging. Curr.
  Opin. Chem. Biol.  \textbf{45},  104--112 (2018)

\end{thebibliography}

\clearpage
\section*{Supplementary Material}

\begin{figure}[h]
	\centering
	\includegraphics[width=0.6\textwidth]{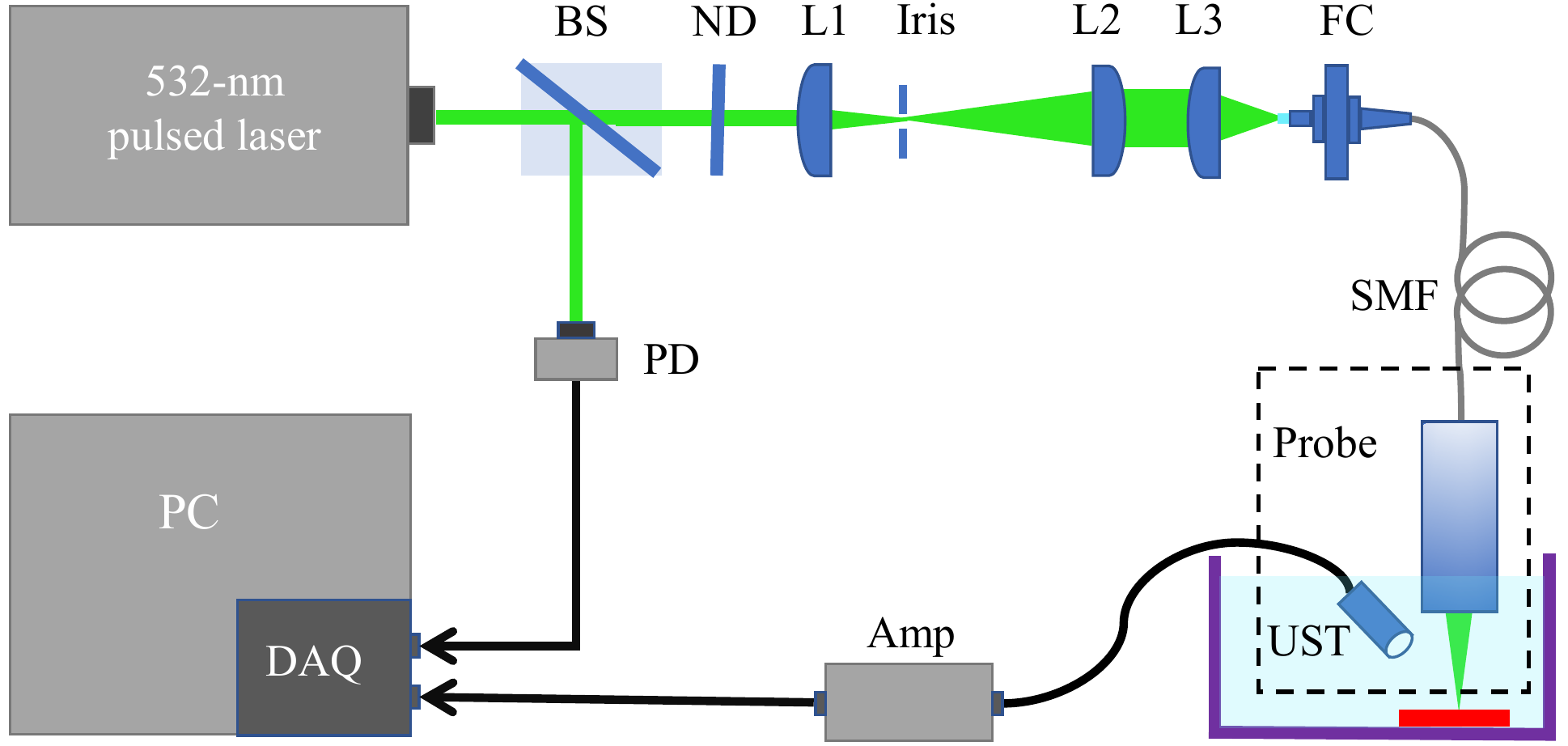}
	\\
	\textbf{Fig. S1.} Schematic of the OR-PAM system. BS, beam splitter; PD, photodiode; ND, neutral density filter; L1, lens \#1; L2, lens \#2; L3, lens \#3; FC, fiber coupler; SMF, signal mode fiber; UST, ultrasound transducer; Amp, preamplifier; DAQ, data acquisition card; PC, personal computer.
	\label{figs1}
\end{figure}	

\begin{figure}[h!]
	\includegraphics[width=\textwidth]{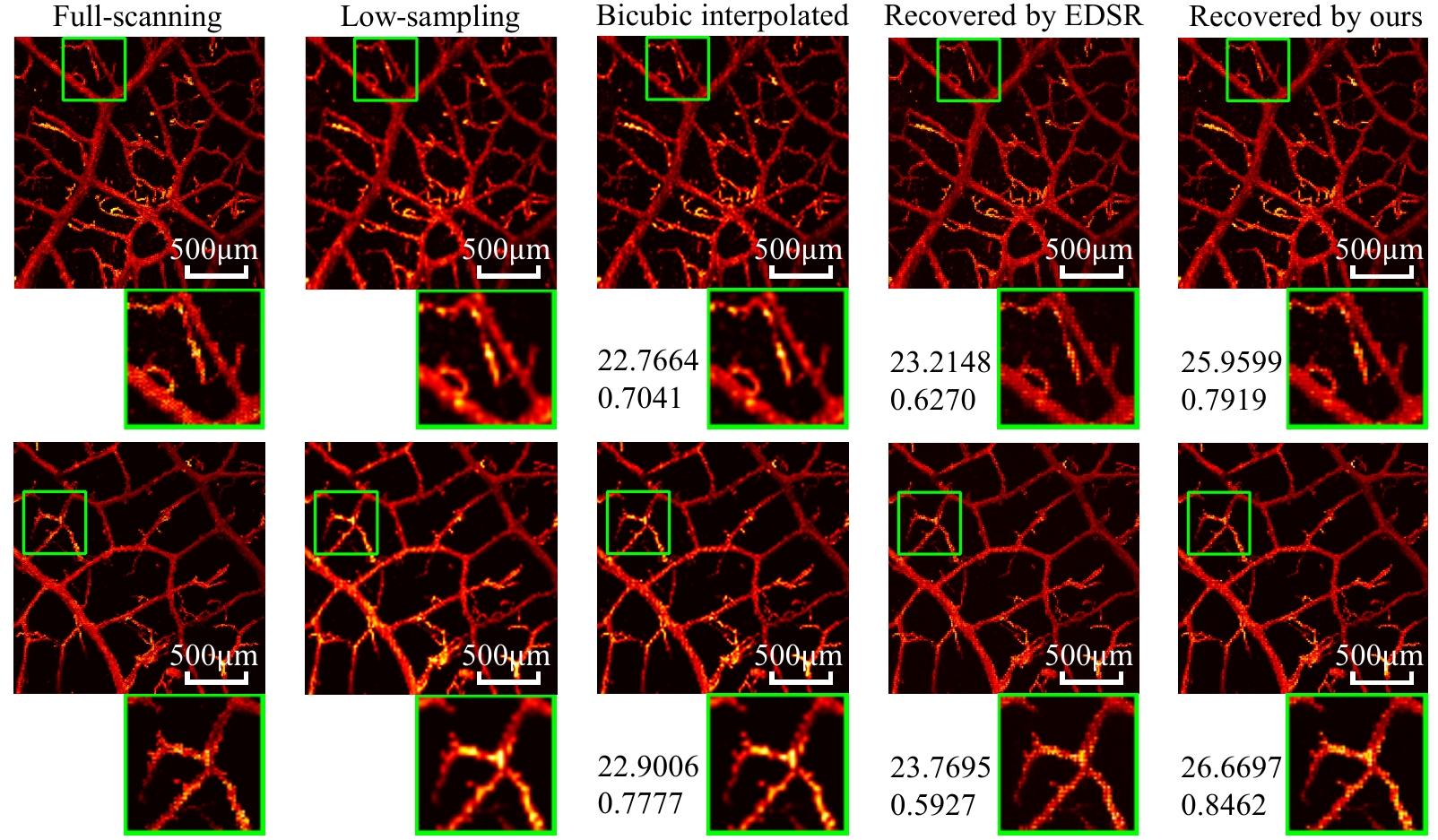}
	\\
	\textbf{Fig. S2.} Example $2\times$ scaling results of the leaf vein experiment. The numbers below images indicate the PSNR and SSIM values compared with the corresponding ground truth. Both samples come from magnolia leaves.
	\label{figs2}
\end{figure}

\begin{figure}[h!]
	\includegraphics[width=\textwidth]{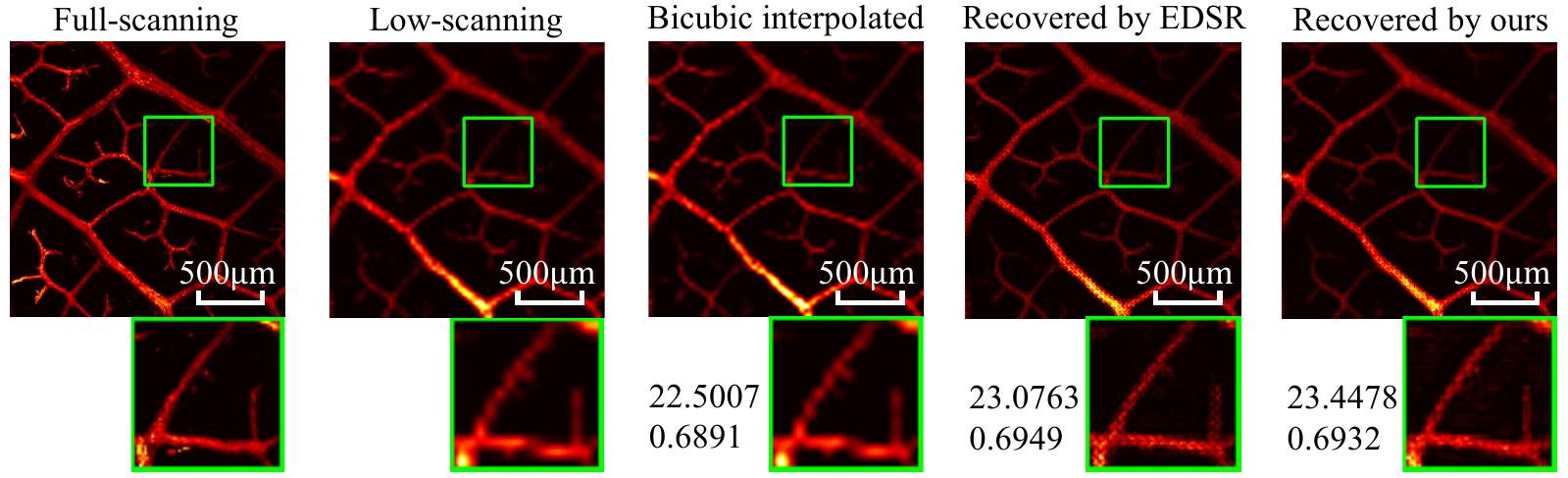}
	\\
	\textbf{Fig. S3.} Experimentally-acquired sparse data verification. The low-scanning image (from an oak leaf) is collected by using large scanning step size experimentally. The numbers below images indicate the PSNR and SSIM values compared with the corresponding full-scanning image. Noise and laser fluctuations might impact the quantitative calculation.
	\label{figs3}
\end{figure}

\begin{figure}[h!]
	\centering
	\includegraphics[width=0.6\textwidth]{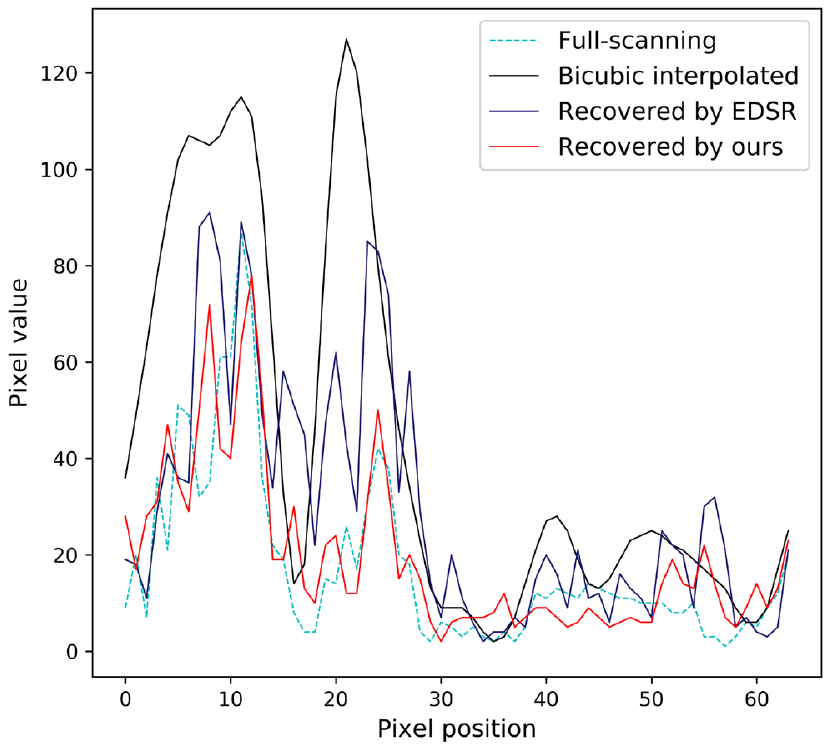}
	\\
	\textbf{Fig. S4.} 1D profiles along the dashed green lines in Fig. 5 of the main text.
	\label{figs4}
\end{figure}

\begin{table}[b]
	\centering
	\textbf{Table S1.} Ablation investigation results of the existence of SE blocks.
	\label{table:s1}
	\begin{tabular}[c]{lcccc}
		\toprule
		\multirow{2}{*}{  }&
		\multicolumn{2}{c}{$2\times$} & \multicolumn{2}{c}{$4\times$} \\
		\cmidrule(lr){2-5}
		& PSNR (dB) & SSIM & PSNR (dB) & SSIM \\
		\midrule
		Without SE blocks & 24.9429 & 0.8124 & 21.7884 & 0.6223\\
		With SE blocks & \textbf{26.1431} & \textbf{0.8183} & \textbf{23.1760} & \textbf{0.7159}\\
		\bottomrule
	\end{tabular}
\end{table}

\end{document}